\documentclass[letters,usenatbib]{mnras}

\usepackage{newtxtext,newtxmath}

\usepackage[T1]{fontenc}

\DeclareRobustCommand{\VAN}[3]{#2}
\let\VANthebibliography\thebibliography
\def\thebibliography{\DeclareRobustCommand{\VAN}[3]{##3}\VANthebibliography}

\usepackage{graphicx}	
\usepackage{amsmath}	
\usepackage{comment}

\newcommand{\masyr}{\hbox{\,mas\,yr$^{-1}$}}
\newcommand{\kms}{\hbox{\,km\,s$^{-1}$}}
\newcommand{\msun}{\hbox{\,M$_\odot$}}

\title[Monoceros OB4]{Monoceros OB4: a new association in Gaia DR2}

\author[P. S. Teixeira et al.]{
P. S. Teixeira,$^{1}$\thanks{E-mail: psdvt@st-andrews.ac.uk}
J. Alves,$^{2}$
A. Sicilia-Aguilar,$^3$
A. Hacar,$^2$
A. Scholz,$^1$
\\
$^{1}$Scottish Universities Physics Alliance (SUPA), School of Physics and Astronomy, University of St. Andrews, North Haugh, Fife, KY16 9SS, St. Andrews, UK\\
$^{2}$University of Vienna,Department of Astrophysics, T\"urkenschanzstrasse 17, A-1180 Vienna, Austria\\
$^3$SUPA, School of Science and Engineering, University of Dundee, Nethergate, Dundee DD1 4HN, UK
}

\date{Accepted XXX. Received YYY; in original form ZZZ}

\pubyear{2015}

\begin{document}
\label{firstpage}
\pagerange{\pageref{firstpage}--\pageref{lastpage}}
\maketitle

\begin{abstract}
We use Gaia DR2 data to survey the classic Monoceros OB1 region and look for the existence of a dispersed young population, co-moving with the cloud complex. An analysis of the distribution of proper motions reveals a 20-30 Myr association of young stars, about 300-400 pc away from the far side of the Mon OB1 complex, along the same general line-of-sight. We characterize the new association, Monoceros OB4, and estimate it contains between 1400 and 2500 stars, assuming a standard IMF, putting it on par in size with NGC\,2264. We find from the internal proper motions that Mon OB4 is unbound and expanding. Our results seem to unveil a larger and more complex Monoceros star formation region, suggesting an elongated arrangement that seems to be at least $300\times60$ pc.

\end{abstract}

\begin{keywords}
ISM: individual objects: Monoceros OB1 cloud complex; Stars: pre-main-sequence; Galaxy: open clusters and association: general; 
\end{keywords}



\section{Introduction}
\label{sec:intro}

Studies of star formation in the Milky Way has been traditionally focused on the densest star forming gas clumps. For the local clouds complexes we have today an essentially complete sample of embedded YSOs with ages of about 1 Myr, mostly found via space-based infrared missions. Almost paradoxically, little is known about the environs of star forming regions, which could contain more evolved YSOs with ages up to 10 Myr, mostly dust free and detectable at optical wavelengths. The existence, or absence, of an extended young optically visible population near active star forming-regions, is critical to establish their star formation history and could inform on the role of feedback in the star formation.
\begin{figure}
    \centering
    \includegraphics[width=\columnwidth]{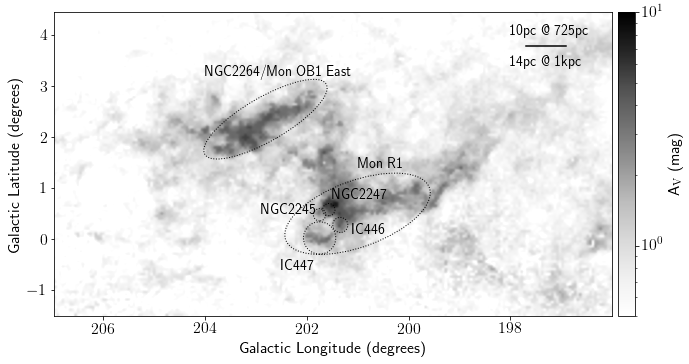}
    \caption{Dust extinction map of the Monoceros OB1 complex, integrated along the line-of-sight up to 1\,kpc \citep{green19}. The most prominent star-forming region is NGC\,2264; it is located within the ellipse representing the approximate area surveyed with \emph{Spitzer} by \citet{rapson14}. Other known star-forming regions, located in Mon R1 region, are marked by dashed circles \citep{bhadari20}. A scale bar is shown for reference.}
    \label{fig:avmap}
\end{figure}
This is the first of a series of papers on the stellar populations in Monoceros OB1, where we conduct a comprehensive wide-field survey to identify and characterize all the co-moving, coeval, and co-distant groups of sources. Monoceros OB1's molecular cloud complex is located between 697\,pc and 784\,pc \citet{zucker19}, and has spawned the NGC\,2264 cluster in the east, and a series of reflection nebula \citep[termed Mon R1 by ][]{bergh66} in the west (see Figure \ref{fig:avmap}).
The most well studied cluster in Mon OB1 is NGC\,2264 \citep[][and references therein]{dahm08},\citep{young06,teixeira06,teixeira07,teixeira08,forbrich10,alencar10,kwon11,teixeira12,marinas13,pineda13,cody14,rapson14,venuti14,mcginnis15,sousa16,tobin15,schneider18,costado18,venuti19,sousa19,mon19a,montil19b,buckner20,jackson20,pearson20}. Mon R1 contains the less well characterized NGC\,2245, NGC\,2247,  IC\,446, and IC\,447 young clusters \citep{glushkov72,dzhakusheva88,casey91,bhadari20}. We report on the discovery of a new cluster, comparable in size to NGC\,2264, that is located behind the molecular clouds.

\section{Gaia DR2 data quality filtering}
\label{sec:data}

We investigated Gaia DR\,2 data \citep{gaiadr218,lindegren18} for the  area shown in Figure \ref{fig:avmap}, corresponding to 
\hbox{196.5\degr\ $<$ l $<$ 206.5\degr} and \hbox{4.5\degr\ $<$ b $<$ -1.5\degr}. These data were cross-matched, via the Gaia DR\,2 source identification number, with the catalog of \citet{bailer-jones18} that provides estimated distances, along with lower and upper bounds of the confidence interval of the estimated distances. \footnote{ Calculating a distance by inverting a parallax is only appropriate when there are no measurement errors. Since there is an error associated with the Gaia DR2 parallaxes, this becomes a non trivial inference problem \citep{BJ15}.}
We filtered the Gaia DR2 data by selecting sources (i) with  parallax/parallax error, $\varpi/\sigma_\varpi$, greater than 10.0 and (ii) with errors in proper motion, $\sigma_{\mu_\alpha^{*}}$ and $\sigma_{\mu_\delta}$, less than 0.2\,mas\,yr$^{-1}$. 

\section{Identification of the new co-moving group and selection of candidate members}
\label{sec:pm}

\begin{figure}
    \centering
    \includegraphics[width=\columnwidth]{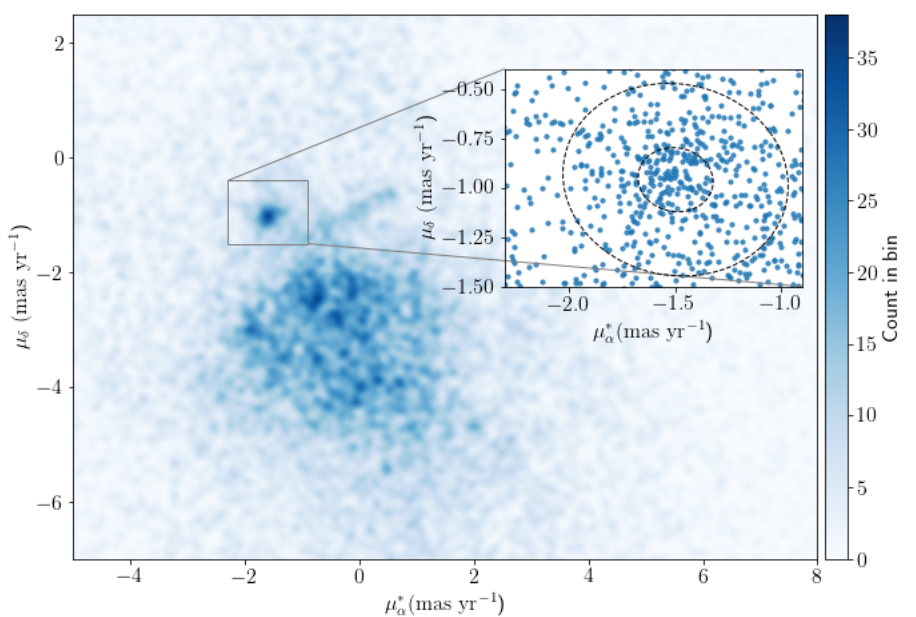}
    \caption{Gaia DR2 proper motion density map for sources within the area shown in Figure \ref{fig:avmap}, that have estimated distances between 1000 and 1300 parsecs, and with the data quality filtering described in Section \ref{sec:data}. The proper motion density colormap has a bin size of 0.08 mas\,yr$^{-1}$, Nyquist sampled.
    The plot shown in the inset contains the proper motions of the sources located within the box on the colormap; the 1- and 3-$\sigma$ contour levels of the fitted bivariate Gaussian are also shown within the inset plot. }
    \label{fig:pm}
\end{figure}

We use a sub-sample of the Gaia DR2 filtered data (as described in Section \ref{sec:data}), consisting of sources that have a lower bound of the estimated distance of 1000\,pc and an upper bound of 1300\,pc  (analysis of other distance layers will be presented in future work). 
Figure \ref{fig:pm} (a) shows the distribution of sources in this sub-sample in the proper motion space; the density map was built using a square bin of size 0.08\masyr, and is Nyquist sampled. The color map shows a well defined density peak, marked by the gray box. The proper motions of the sources that fall within the box are shown in the inset, where they are fit using robust parameter estimation of a bivariate Gaussian \citep{astroML, astroMLText}; we retain sources whose proper motion fall within the 3-$\sigma$ contour level of the ellipse that is centered on \hbox{($\mu_\alpha^{*},\mu_\delta$)=(-1.5, -0.9)} \masyr. 

\begin{figure*}
    \centering
    \includegraphics[width=0.9\textwidth]{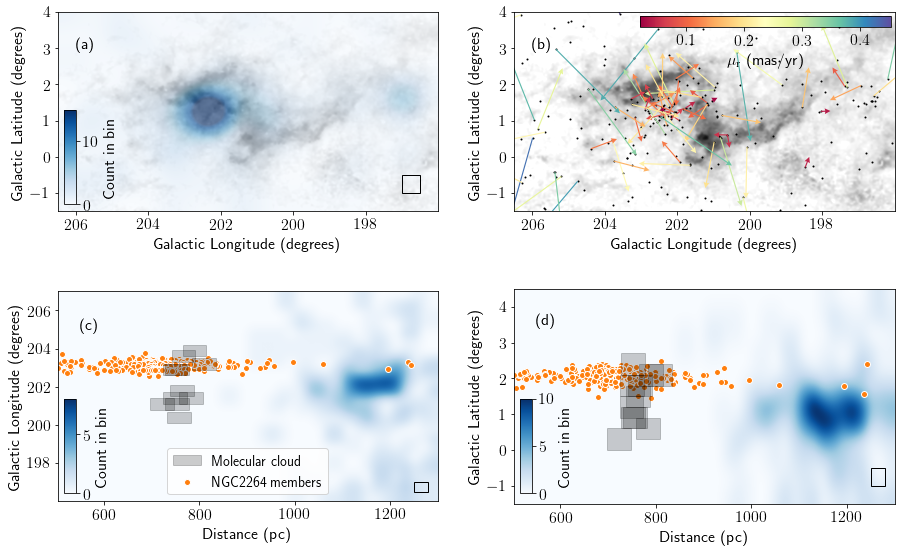}
    \caption{3D spatial representation of the new association Mon  OB4. Panel (a) shows the $(l.b)$ spatial density (blue colorscale) overplotted on the dust map (c.f. Figure \ref{fig:avmap}). Panel (b) shows} the $(l.b)$ spatial distribution of the candidate members ( black points) overplotted on the dust map, with the vectors representing their proper motion relative to the mean of Mon OB4. Panels (c) and (d) show the Mon OB4 density distribution of $l$ and $b$, respectively, as a function of distance (blue colormap).  Also represented are the Mon OB1 molecular clouds (gray boxes) and NGC\,2264 members \citep{venuti14} (orange points). The distances to the clouds are taken from \citet{zucker19}; cloud depths are as yet unknown and we take a value of 50\,pc for schematic purposes. The bottom left rectangle in panels (a), (c), and (d) correspond to the pixel size of the respective density map (20\,pc for distance and 0.5\degr\ for $l$ and $b$).
    \label{fig:3D}
\end{figure*}

We used the same sub-sample of data to build a density map in the transverse velocity phase space. The transverse velocity along right ascension, $V_\mathrm{T_\alpha}$, and declination, $V_\mathrm{T_\delta}$, are given by:
\begin{align}
& V_\mathrm{T_\alpha}^* = 4.74 \mu_\alpha^{*} d = 4.74 \mu_\alpha cos(\delta) d, \\
& V_\mathrm{T_\delta} = 4.74 \mu_\delta d,
 \end{align}
\noindent where $d$ corresponds to the estimated distance to the source given in pc, the proper motions are given in \arcsec yr$^{-1}$, and the transverse velocities are calculated in \kms. 
The  transverse velocity density map was built using a square bin of size 0.37\kms, Nyquist sampled. We retained sources within the 3-$\sigma$ contour level of the bivariate Gaussian, in the respective proper motion and transverse velocities phase spaces, as a first selection criteria for members of the new co-moving group. Table \ref{tab:kin} summarizes the parameters of the 3-$\sigma$ ellipses, including the rotation angles for the proper motion ($\theta_\mu$) and transverse velocity ($\theta_{V_T}$). We selected sources that satisfied both $\mu$ and $V_\mathrm{T}$ criteria simultaneously, in order to minimize contamination. We expanded the search for candidate members, with the above kinematic criteria, for distances down to 400\,pc, in order to compare with the known population of Mon OB1 and to account for the large distance uncertainties. The new co-moving group has 242 candidate members (listed in Table \ref{tab:sources}), and is located at a mean distance of 1182 $\pm$122\,pc.
We show in panel (a) of Figure \ref{fig:3D} the spatial distribution of the co-moving group in comparison to the molecular clouds in the Mon OB1 complex.
We calculate the median values of coordinates of the candidate members and find the center of the co-moving group to be $(l,b)$=(202.2\degr,1.1\degr), with an estimated radius of 0.5\degr. We should note that since the co-moving group is located behind the molecular cloud complex, we are missing those members that are too extincted to be detected by Gaia, and as such the actual co-moving group center may be shifted to the west.

\begin{table}
    \caption{Kinematic selection criteria.}
    \centering
    \begin{tabular}{c|c}
        \hline
        \hline
        $\mu_\alpha^{*} \pm 3\sigma_{\mu_\alpha^{*}}$ =  -1.52 $\pm$ 0.47 \masyr & 
        $V_\mathrm{T_\alpha}^* \pm 3\sigma_{V_\mathrm{T_\alpha}^*}$ = -6.47 $\pm$ 1.23 \kms  \\
        $\mu_\delta \pm 3\sigma_{\mu_\delta}$ = -0.96 $\pm$ 0.39 \masyr & $V_{\mathrm{T}_\delta} \pm 3\sigma_{V_\mathrm{T_\delta}}$ = -3.80 $\pm$ 0.99 \kms\\ 
        $\theta_{\mu}$ = -0.24\degr & 
        $\theta_{V_\mathrm{T}}$ = 1.35\degr\\
        \hline
    \end{tabular}
    \label{tab:kin}
\end{table}

\begin{table}
\caption{List of candidate members of the new cluster. The full catalog will be available online.}
\label{tab:sources}
\begin{tabular}{lcccc}
\hline
\hline
Gaia DR2 Source ID & l (deg) & b (deg) & Dist. (pc) & Av (mag) \\
\hline
3369260560768525056 & 196.187 & 2.147 & 1076 & 0.6 \\ 
3357029833938241024 & 197.528 & 3.583 & 1165 & 0.6 \\ 
3356941666849693568 & 197.588 & 2.670 & 1275 & 1.1 \\ 
3356860921465018240 & 196.439 & 1.815 & 1201 & 0.6 \\ 
3356834975564998784 & 196.489 & 1.639 & 1043 & 0.7 \\ 
\hline
\end{tabular}
\end{table}

\section{Age and mass function}
\label{sec:kinematics}

\begin{figure}
    \centering
    \includegraphics[width=0.9\columnwidth]{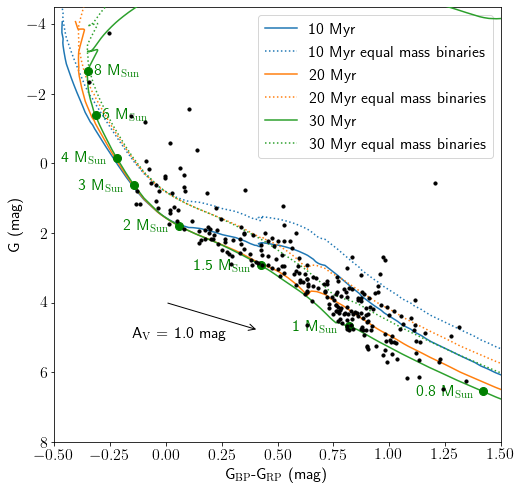}
    \caption{Color-magnitude diagram for the sources in the co-moving group,  with overplotted MIST isochrones \citep{choi16,dotter16} for single stars (solid lines) and equal-mass binaries (dotted lines) for 30, 20, and 10\,Myr (green, orange, and blue, respectively)} . 
    \label{fig:cmd}
\end{figure}

Figure \ref{fig:cmd} shows the de-reddened Gaia DR2 color-magnitude diagram. We use the Bayestar dust extinction map \citep{green19}, to estimate the individual extinction for each source. 
According to this method, the sources have an average extinction of A$_\mathrm{V} = {1.0} \pm 0.5$\,mag. We determine an approximate age between 20 and 30\,Myr. Spectroscopic follow-up observations are needed to obtain spectral types and more exact extinction measurements to determine a more precise age for the new group. We note that several other factors also contribute to a spread in the distribution of the sources in the color-magnitude diagram, such as binarity, distance uncertainties, and stellar variability \citep{jeffries11}.

\begin{figure}
    \centering
    \includegraphics[width=0.9\columnwidth]{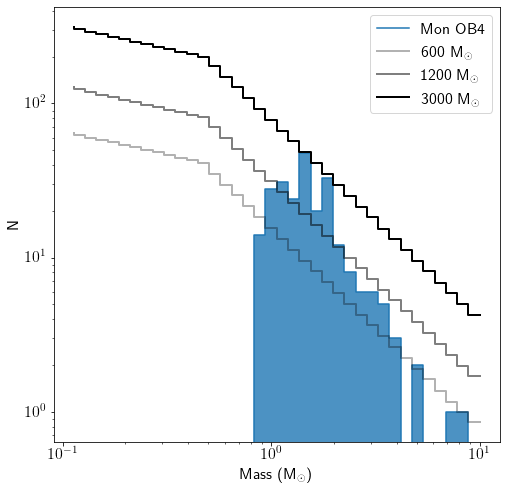}
    \caption{ Present day mass function of the Mon OB4 (solid histogram). For comparison, we over plot the initial mass functions of simulated clusters of total masses ranging between  600\msun\ and  3000\msun\ (open histograms).}
    \label{fig:pdmf}
\end{figure}

Using the de-reddened color-magnitude diagram, and the MIST mass tracks \citep{choi16,dotter16}, we estimate individual stellar masses and built the mass function shown in Figure \ref{fig:pdmf}. In order to estimate the total mass of the new co-moving group, we 
generated clusters of different total masses, using the \citet{kroupa01} initial mass function (IMF), and the publicly available python package \emph{imf}\footnote{available at \url{https://github.com/keflavich/imf}}. We show in Figure \ref{fig:pdmf}\  the IMFs of these simulated clusters, with total masses ranging from 600\msun\ to  3000\msun, and show that the total mass of the new co-moving group is estimated to be between  600 and 1200\msun\ (containing between 1400 and 2500 stars). We remind the reader that, since the new group is located behind the molecular cloud complex, we are missing those members that are too extincted to be detected by Gaia DR2. For comparison, regarding the population of NGC\,2264, \citet{teixeira12} calculated a total number of sources of 1436 $\pm$ 242, while \citep{buckner20} estimate a population of 1795 (differences in the areas surveyed and the wavelength coverage may account for the slightly differing values). The new co-moving group is therefore comparable in population size to NGC\,2264. We henceforth refer to the new co-moving group as the Monoceros OB4 association.

\section{Discussion}
\subsection{2D kinematic structure: expansion}
\label{sec:kinematics}

 We transformed the proper motions from the ICRS ($\alpha, \delta$) plane to the Galactic ($l, b$) plane, and subtracted the mean proper motion values of Mon OB4, ($\mu_l, \mu_b$) = (0.09, -1.80)\masyr\ from the individual sources in order to ascertain their relative motions. Panel (b) of Figure \ref{fig:3D} shows these relative proper motion vectors, color-coded by their value. To simplify the diagram, we show only the vectors for the brightest sources, with G $<$ 2.5\,mag. As can be seen, the proper motion value is correlated with the distance of the source to the center of Mon OB4: sources located closest to the center have lower proper motions, while sources located further away have larger proper motions. This is a classic signature of an expanding cluster, also observed in $\lambda$ Orionis \citep{kounkel18}, and modeled by  \citet{arnold19}.  We do note however, that due to the lack of radial velocity information we are unable to correct for virtual expansion \citep{brown97}.

The virial mass of a cluster is given by \citep{mihalas68}:
\begin{equation}
    M = 466 <R> <v^2>,
\end{equation}
\noindent where $<R>$ is the mean distance between the stars in pc, $<v^2>$ is the square of the dispersion of space velocities in \kms. To calculate the virial mass of the cluster, we create a sub-sample corresponding to the region where the stellar density is greater than 5-$\sigma$, which corresponds to the following box: 202.53\degr\ $< l <$ 201.78\degr, 1.39\degr\ $< b <$ 0.74\degr.
For this sub-sample, $<R>$ = 4.4\,pc and $<v^2>$ = 1.1\kms\ (in the absence of radial velocities we assume isotropy). We calculate a virial mass of 2364\msun,  which is greater than the total estimated mass of Mon OB4 association (by a factor of two, at least). This is a first rough estimate for the current data (keeping in mind the aforementioned caveats such as missing sources that are too extincted for Gaia DR2 detection, velocity isotropy),  suggesting that Mon OB4 is unbound, which is in agreement with the expansion signature described above and seen in Figure \ref{fig:3D} panel (a).

\subsection{3D spatial distribution}

We carried out an analysis of the maximum likelihood in the 3D multivariate Gaussian space of parallax and proper motion \citep{lindegren00, roccatagliata20}, to explore whether sources located at $l <$ 200\degr\ (farthest from the center of Mon OB4, on the plane-of-the-sky) could be part of a different population, but we found no statistically significant difference. The population is consistent with a single peak in the parallax and proper motion space.
Figure \ref{fig:3D} shows the distribution of Mon OB4 in  Galactic Longitude and Latitude coordinates as a function of distance in panels  (c) and (d), respectively. Also shown for comparison purposes are the clouds \citep[distances taken from][]{zucker19} and the NGC\,2264 cluster \citep[spectroscopically confirmed members taken from][]{venuti14}.
At first glance, the distance between the new Mon OB4 association and the Mon OB1 star-forming clouds, $\sim$400\,pc, seems too large for these two structures to be physically connected. \citet{stassun18} found a systematic offset in the parallaxes reported in Gaia DR2 of -82 $\pm$ 33\,$\mu$as. 
At a distance of 1183\,pc, this corresponds to an average distance overestimation of 105\,pc bringing the distance of the new  association Mon OB4 to 1078\,pc, and reducing the separation to the clouds to $\sim$300\,pc. A careful Gaia DR2 analysis of the Taurus star-forming region has revealed that the complex has a depth of 200\,pc \citep{roccatagliata20}. Presumably more massive regions would span larger depths, so it is not entirely unfeasible that Mon OB4 could be associated with the star-forming complex in Mon OB1. For example, \citet{Chen2019-fi}, also using Gaia DR2 data, found that the young stellar population in the large Orion star forming region extends for about 300\,pc, with ages spanning 0-20 Myr, similar to our results in Mon OB1. 
Finally, \cite{Alves2020-um} found that the location of the large majority of the local star forming regions is not random but highly correlated. In particular, they found an undulating 2.7 kpc structure, the Radcliffe Wave, likely associated with the Local Arm of the Galaxy, containing all the major local star forming regions from Canis Major to Cygnus. Monoceros OB1 is not part of the Radcliffe Wave (see their interactive Figure 2\footnote{\href{https://faun.rc.fas.harvard.edu/czucker/Paper\_Figures/radwave.html}{https://faun.rc.fas.harvard.edu/czucker/Paper\_Figures/radwave.html}}), but the extended configuration of Mon OB1 proposed here is not surprising in the context of the spatial distribution of star formation in the Local Neighborhood. It is suggestive, on the other hand, of a possible connection between Mon OB1 and the Rosette nebula, at 1.6 kpc  \citep[c.f. Cepheus Spur,][]{pantaleoni21}.
Although it is likely that the new population presented in this Letter is associated with the star formation event that formed NGC\,2264, radial velocity data are needed to further investigate and confirm this scenario, by determining how the new  Mon OB4 association and clusters emerging from the clouds (e.g. NGC\,2264 and IC\,447) are moving with respect to each other in 3D.

\section{Summary of conclusions}
\label{sec:summary}

We report on the discovery of a new  co-moving group of young sources in Monoceros, also independently identified by \citet{kounkel19,kounkel20}.  We name the new group as the Monoceros OB4 association; it is comparable in estimated population size to NGC\,2264, and with the following characteristics:

\begin{itemize}
    \item its central coordinates are $(l,b)$ = (202.2\degr, 1.1\degr);
    \item its mean proper motion is given by ($\mu_{\alpha}^*,\mu_\delta$) = (-1.52, -09.6)\,\masyr;
    \item it is located at a distance of $\sim$1\,kpc, placing it behind the molecular clouds associated with Mon OB1 East (NGC\,2264) and Mon OB1 West (NGC\,2245, NGC\,2247, IC\,446 and IC\,447);
    \item its age is  estimated to be between 20 and 30\,Myr;
    \item it has a  total mass estimate between  600 and 1200\msun and  between 1400 and 2500 sources;
    \item it is unbound and appears to be expanding;
    \item Our results seem to unveil a larger and more complex Monoceros star formation region, suggesting an elongated arrangement that seems to be at least $300\times60$ pc.
\end{itemize}

Radial velocity observations are needed to further investigate the motion of the new population in 3D with respect to the clusters associated with the molecular cloud complex.
\section*{Acknowledgements}

 We thank our anonymous referee for a constructive report. We thank S. Meingast and V. Roccatagliata for helpful discussions. This project was supported by STFC grant ST/R000824/1.  ASA is partly supported by STFC grant ST/S000399/1. This project has received funding from the European Research Council (ERC) under the European Union’s Horizon 2020 research and innovation programme (Grant agreement No. 851435). This  work  has  made  use  of data  from  the  European  Space  Agency  (ESA)  mission Gaia (https://www.cosmos.esa.int/gaia),  processed  by  the Gaia Data  Processing  and  Analysis Consortium (DPAC, https://www.cosmos.esa.int/web/gaia/dpac/consortium). Funding for the DPAC has been provided by national institutions, in particular the institutions participating in the Gaia Multilateral Agreement. 
This research has made use of Python, https://www.python.org, NumPy \citep{vanderwalt11}, and Matplotlib \citep{hunter07}.
This research made use of APLpy, an open-source plotting package for Python and hosted at  http://aplpy.github.com \citep{robitaille12}. This research made use of Astropy, a community-developed core Python package for Astronomy \citep{robitaille13}. This  research  made  use  of  TOPCAT,  an  interactive  graphical viewer  and  editor  for  tabular  data  \citep{taylor05}.  

\section{Data availability}

The data underlying this article will be made available in \url{http://cdsarc.u-strasbg.fr/}


\bibliographystyle{mnras}
\input{mnras.bbl}



\bsp	
\label{lastpage}
\end{document}